\documentclass[11pt]{article}
\usepackage[margin=3cm]{geometry}
\usepackage{t1enc}
\usepackage{graphicx}
\usepackage{mathptmx}

\usepackage{graphicx}
\usepackage{dcolumn}
\usepackage{bm}
\DeclareGraphicsExtensions{.eps,.EPS}

\begin{document}

\title{Maximum Entropy distributions of correlated variables with prespecified
 marginals}

\author{Hern\'an Larralde \\
        Instituto de Ciencias F\'\i sicas, UNAM. Apdo. Postal 48-3.\\
        C.P. 62251, Cuernavaca, Morelos, M\'exico.}%
\date{\today}

\maketitle

\begin{abstract}
The problem of determining the joint probability distributions for
correlated random variables with pre-specified marginals is
considered. When the joint distribution satisfying all the required
conditions is not unique, the ``most unbiased'' choice corresponds to
the distribution of maximum entropy. The calculation of the maximum
entropy distribution requires the solution of rather complicated
nonlinear coupled integral equations, exact solutions to which are
obtained for the case of Gaussian marginals; otherwise, the solution
can be expressed as a perturbation around the product of the marginals
if the marginal moments exist.
\end{abstract}




Consider the situation in which we are given two random variables,
say $X_1\in I_1$ and $X_2\in I_2$, which we know to be distributed as
$P_1(X_1)$ and $P_2(X_2)$, respectively. Further, assume we know the
variables to be correlated; for example, assume we are given the
covariance $\Gamma_{12}=\langle X_1X_2\rangle -\langle X_1\rangle \langle X_2\rangle  \neq 0$.  We are now required to construct the joint probability distribution $P_{(1,2)}(X_1,X_2)$ with the prescribed marginals, $P_1(X_1)$ and
$P_2(X_2)$, and covariance $\Gamma_{12}$.

This and other similar problems arise in a wide variety of contexts,
ranging from the description of correlated financial instruments in
economics \cite{Bouchaud}, EEG signals \cite{Markus,Schindler} in
medicine, to systems out of equilibrium in statistical mechanics
\cite{Nicolis,Dorfman,Zarate,HL}; to name but very few. Actually, in
finance and other fields of intense applied statistics
\cite{Schmidt,Genest,Piant1,Piant2}, it has become popular to describe
interdependent random variables with given marginals using
``copulas''. The idea there is that the ``interdependence'' of, say, N
random variables described by cumulative marginal distributions
$F_i(X_i)$ is encoded in the N-dimensional cumulative distribution
function with uniform marginals, the copula: $C(u_1,...,u_N): [0,1]^N
\rightarrow [0,1]$, with $C(1,1...u_j...,1)=u_j$. The complete
description is achieved through the joint cumulative distribution
$F(X_1,...,X_N)=C(F_1(X_1),...,F_N(X_N))$. This approach treats the
individual statistics of the random variables, the marginals,
separately from the interdependence of said variables, allowing, for
example, to change the marginals keeping the interdependence, the
copula, fixed. These tools are extremely powerful and general, but
hard to estimate directly from data. Another caracterization of data
interdependence relates to whether there is a {\it causal} relation
between the variables, which can be tested along the lines originally
proposed by Granger in the context of econometrics \cite{granger}. In
contrast to these more sophisticated methods, and while far from a
complete description of the interdependence structure of much data,
the correlation between two random variables is a frequently and
easily measurable quantity, extensively used in many disciplines,
including, of course, physics.

As it stands, however, the problem may be ill posed since there could
be infinitely many distributions (or none at all) satisfying the
conditions of the prespecified marginals and given covariance. To lift
the ambiguity, when it arises, we follow Jaynes \cite{Jaynes} and
require the joint distribution function to be that which maximizes the
relative entropy or, equivalently, minimizes the discrimination
information over the product of the marginals. This choice is, as
argued by Jaynes, the ``least biased'' distribution which is
consistent with the restrictions: ``the maximization of entropy is
[...] a method of reasoning which ensures that no unconscious
assumptions have been introduced.''\cite{Jaynes} 
To this point, then,
the problem is formally straightforward: we need to find the extreme
of the entropy functional subject to the appropriate
restrictions. That is, we require $P_{(1,2)}(x,y)$ such that
\begin{eqnarray}
0&=&\delta\left[~~ \int\limits_{I_1\times I_2} P_{(1,2)}(X_1,X_2)\ln
\left(\frac{P_{(1,2)}(X_1,X_2)}{P_1(X_1)P_2(X_2)}\right)dX_1 dX_2 + \right.
\lambda_{12} \int\limits_{I_1\times I_2} X_1X_2 P_{(1,2)}(X_1,X_2) dX_1 dX_2
\cr &&
\cr && +\left. \int\limits_{I_1\times I_2} \left(a(X_1)+b(X_2)\right)P_{(1,2)}(X_1,X_2)dX_1 dX_2\right].
\label{MinS}
\end{eqnarray}
(we do not need to condition the distribution to be normalized, as the
marginals are assumed to be already normalized). The required
distribution can be written as
\begin{equation}
 P_{(1,2)}(X_1,X_2)= P_1(X_1)P_2(X_2)e^{-a(X_1)-b(X_2)-\lambda_{12} X_1X_2 -1}\equiv
 P_1(X_1)P_2(X_2){\mathcal A}(X_1){\mathcal B}(X_2)e^{-\lambda_{12} X_1X_2}.
\label{Form}
\end{equation}
The Lagrange multipliers $a(X_1)$, $b(X_2)$ (or equivalently, the
functions ${\mathcal A}(X_1)$ and ${\mathcal B}(X_2)$), and the
constant $\lambda_{12}$, are chosen to enforce the restrictions, which
result in the set of coupled nonlinear integral equations
\begin{eqnarray}
1={\mathcal A}(X_1)\int\limits_{I_2}P_2(X_2){\mathcal
  B}(X_2)e^{-\lambda_{12} X_1X_2} dX_2, \qquad X_1 \in I_1 \cr
1={\mathcal B}(X_2)\int\limits_{I_1}P_1(X_1){\mathcal
  A}(X_1)e^{-\lambda_{12} X_1X_2} dX_1 \qquad X_2 \in I_2
\label{Coupledeqs}
\end{eqnarray}
plus a condition on the value of $\lambda_{12}$:
\begin{equation}
 \int\limits_{I_1\times I_2} X_1X_2 P_1(X_1)P_2(X_2){\mathcal A}(X_1)
{\mathcal B}(X_2)e^{-\lambda_{12} X_1X_2}dX_1 dX_2
=\Gamma_{12} + \langle X_1\rangle \langle X_2\rangle 
\label{lambda}
\end{equation}
where the mean values $\langle X_1\rangle $ and $\langle X_2\rangle $
are calculated from the corresponding marginals. The above equations
may be rewritten in the slightly more compact form
\begin{eqnarray}
P_1(X_1)= Q_1^{(2)}(X_1)\int\limits_{I_2} Q_2^{(2)}(X_2)e^{-\lambda_{12}^{(2)} X_1X_2} 
dX_2, \qquad&& X_1 \in I_1 \cr
P_2(X_2)= Q_2^{(2)}(X_2)\int\limits_{I_1} Q_1^{(2)}(X_1)e^{-\lambda_{12}^{(2)} X_1X_2}
dX_1  \qquad&& X_2 \in I_2 
\label{Coupledeqs1}
\end{eqnarray}
and
\begin{equation}
 \int\limits_{I_1\times I_2} X_1X_2 Q_1^{(2)}(X_1)
 Q_2^{(2)}(X_2)e^{-\lambda_{12}^{(2)} X_1X_2}dX_1 dX_2 =\Gamma_{12}
\label{lambda1}
\end{equation}
where $Q_1^{(2)}(X_1)=P_1(X_1){\mathcal A}(X_1)$ and
$Q_2^{(2)}(X_2)=P_2(X_2){\mathcal B}(X_2)$. The superscripts have been added to
indicate that these quantities are elements of the joint distribution
of two variables. 

Though far from opening the way to a full solution, it is worthwhile
noting that equations (\ref{Coupledeqs1}) can be decoupled by
multiplying the first one, say, by $e^{-\lambda_{12}^{(2)} X_1 Y}$ and
integrating over $X_1$. Then
\begin{eqnarray}
{\tilde P}_1(\lambda_{12}^{(2)} Y)&\equiv& \int\limits_{I_1} P_1(X_1) 
e^{-\lambda_{12} X_1 Y}dX_1 =
\int\limits_{I_2} Q_2^{(2)}(X_2) \int\limits_{I_1} Q_1^{(2)}(X_1)
e^{-\lambda_{12}^{(2)} (X_2+Y) X_1 } dX_1 dX_2\cr && \cr
&=&\int\limits_{I_2}\frac{ Q_2^{(2)}(X_2) P_2(X_2+Y)}{Q_2^{(2)}(X_2+Y)} dX_2,
\end{eqnarray}
where ${\tilde P}_1(Y)$ is the Laplace transform of $P_1(X)$. This is
a rather difficult nonlinear integral equation that determines
$Q_2^{(2)}(X_2)$, up to a multiplicative constant, in terms of the
marginals and the covariance.

If the variables are discrete rather than continuous, similar
expresions are obtained with summations instead of integrals. Either
way, the above equations turn out to be extremely hard to solve for
arbitrary marginals.

The generalization to more than two variables, while formally equally
simple, gives rise to a rather interesting situation. To illustrate
this, consider the case of three variables: $X_1,X_2$ and $X_3$ with
their respective marginals $P_1(X_1),P_2(X_2)$ and $P_3(X_3)$; and
covariance matrix elements $\Gamma_{12},\Gamma_{23}$ and $\Gamma_{13}$ (in
passing, note that only the off diagonal components of the covariance
matrix can be introduced as constrains, the diagonal elements are
fixed by the marginals). If we follow the procedure outlined above for
two variables,-maximizing entropy relative to the product of the
marginals, constrained to the appropriate marginals and correlations-,
it is easy to see that the joint probability distribution should be of
the following form:
\begin{equation}
P_{(1,2,3)}(X_1,X_2,X_3)=Q_1^{(3)}(X_1)Q_2^{(3)}(X_2)Q_3^{(3)}(X_3)
e^{-\lambda_{12}^{(3)} X_1X_2 - \lambda_{23}^{(3)}X_2X_3 -
  \lambda_{13}^{(3)}X_1X_3}
\label{3point1}
\end{equation}
where the functions $Q_i^{(3)}(X_1)$ are related to the Lagrange multipliers
that constrain the marginals:
\begin{equation}
P_1(X_1)=Q_1^{(3)}(X_1)\int \limits_{I_2\times I_3}Q_2^{(3)}(X_2)Q_3^{(3)}(X_3) e^{-\lambda_{12}^{(3)} X_1X_2
 - \lambda_{23}^{(3)}X_2X_3 - \lambda_{13}^{(3)}X_1X_3} dX_2 dX_3
\label{margin2}
\end{equation}
and so on. 

Now the question arises as to whether it must also be true that the
distribution obtained by integrating $P_{(1,2,3)}(X_1,X_2,X_3)$ over
$X_3$, say, must have the form of the constrained maximum entropy
joint distribution for two variables discussed above. That is,
whether:
\begin{eqnarray}
\int\limits_{I_3}P_{(1,2,3)}(X_1,X_2,X_3)dX_3 &=& Q_1^{(3)}(X_1) Q_2^{(3)}(X_2) e^{-\lambda_{12}^{(3)} X_1X_2}\int
\limits_{I_3}Q_3^{(3)}(X_3) e^{ - (\lambda_{23}^{(3)}X_2 + \lambda_{13}^{(3)}X_1)X_3} dX_3 
\cr
&=& Q_1^{(2)}(X_1) Q_2^{(2)}(X_2)e^{-\lambda_{12}^{(2)} X_1X_2} = P_{(1,2)}(X_1,X_2).
\label{condition}
\end{eqnarray}
If so, this would in turn imply that, independently of what the marginals
$P_i(X_i)$ are, the integral that appears above can always be resolved as
\begin{equation}
\int \limits_{I_3}Q_3^{(3)}(X_3) e^{ - (\lambda_{23}^{(3)}X_2 + \lambda_{13}^{(3)}X_1)X_3} dX_3 = q_1^{(3)}(X_1) q_2^{(3)}(X_2)e^{-\mu_{12}^{(3)} X_1X_2},
\label{condition1}
\end{equation}
in terms of which we can express $Q_1^{(2)}(X_1)=q_1^{(3)}(X_1)
Q_1^{(3)}(X_1)$, $Q_2^{(2)}(X_2)=q_2^{(3)}(X_2) Q_2^{(3)}(X_2)$ and
$\lambda_{12}^{(2)}=\lambda_{12}^{(3)}+\mu_{12}^{(3)}$.  

While it turns out that for the simple cases considered in this paper,
eq .(\ref{condition}) is indeed satisfied, I cannot find any reason why
it should be true in general. Actually, let me consider another joint
distribution $\Pi_{(1,2,3)}(X_1,X_2,X_3)$ defined to be the maximum
entropy distribution conditioned so that integrating over any variable
yields the corresponding two point maximum entropy distributions
discussed above:
\begin{eqnarray}
&&0=\delta\left[~~ \int\limits_{I_1\times I_2\times I_3} \Pi_{(1,2,3)}(X_1,X_2,X_3)\ln
\left(\frac{\Pi_{(1,2,3)}(X_1,X_2,X_3)}{p_1(X_1)p_2(X_2)p_3(X_3)}\right)dX_1 dX_2 dX_3 + 
\right.
\cr &&
\cr &&+\left. \int\limits_{I_1\times I_2\times I_3} \left(\alpha_{(1,2)}(X_1,X_2)+\alpha_{(2,3)}(X_2,X_3)
+\alpha_{(1,3)}(X_1,X_3)\right)\Pi_{(1,2,3)}(X_1,X_2,X_3)dX_1 dX_2 dX_3\right]
\nonumber
\label{MinSPi}
\end{eqnarray}
the solution to which can be written as
\begin{equation}
\Pi_{(1,2,3)}(X_1,X_2,X_3)=F_{(1,2)}(X_1,X_2)F_{(2,3)}(X_2,X_3)F_{(1,3)}(X_1,X_3)
\label{altern}
\end{equation}
where the functions $F_{(i,j)}(X_i,X_j)$, are simply related to the
Lagrange multipliers $\alpha_{(i,j)}(X_i,X_j)$ and satisfy nonlinear
integral equations that enforce the conditions imposed on
$\Pi_{(1,2,3)}(X_1,X_2,X_3)$, namely
\begin{equation}
P_{(1,2)}(X_1,X_2)=F_{(1,2)}(X_1,X_2)\int\limits_{I_3}F_{(2,3)}(X_2,X_3)
F_{(1,3)}(X_1,X_3)dX_3
\end{equation}
and so on. The conditions on the distributions $P_{(i,j)}(X_i,X_j)$
ensure that $\Pi_{(1,2,3)}(X_1,X_2,X_3)$ has the correct 1-point
marginals $P_i(X_i)$ for $i=1,2,3$, and covariance
$\Gamma_{i,j}$. It should be noted that $P_{(1,2,3)}(X_1,X_2,X_3)$, as
expressed in eq. (\ref{3point1}), can be written in the form shown in
eq. (\ref{altern}).  However, the conditions on
$\Pi_{(1,2,3)}(X_1,X_2,X_3)$ are as restrictive or more than those on
$P_{(1,2,3)}(X_1,X_2,X_3)$, thus, it should not necesarily be the case
that $P_{(1,2,3)}(X_1,X_2,X_3)=\Pi_{(1,2,3)}(X_1,X_2,X_3)$. Conversely, it
does not appear to be necessarily true that
\begin{equation}
P_{(1,2)}(X_1,X_2)=\int\limits_{I_3}P_{(1,2,3)}(X_1,X_2,X_3) dX_3
\label{2point3}
\end{equation}
etc., so that one could end with the somewhat uncomfortable situation
in which the two point marginals obtained from the maximum entropy
three point distribution may themselves not be maximum entropy two
point distributions.

We now turn to simple cases for which the required maximum entropy
distributions can be calculated explicitly. First, however, for the
trivial case of ``uncorrelated'' variables (i.e. the case in which
$\Gamma_{i,j}=0$), the required maximum entropy joint distribution is
indeed the product of the marginals.  The first nontrivial example is
the case of two correlated random variables with Gaussian marginal
distributions, say:
\begin{equation}
P_1(X_1)=\sqrt{\frac{\alpha}{2\pi}} e^{-\alpha x_1^2/2},
\end{equation}
\begin{equation}
P_2(X_2)=\sqrt{\frac{\beta}{2\pi}} e^{-\beta x_2^2/2}
\end{equation}
and let the correlation parameter be $\Gamma=\langle x_1x_2\rangle $.  
Then, to determine the joint distribution we need to solve
\begin{equation}
\sqrt{\frac{\alpha}{2\pi}} e^{-\alpha
  X_1^2/2}=Q_1(X_1)\int\limits_{-\infty}^\infty Q_2(X_2) e^{-\lambda
  X_1X_2} dX_2
\end{equation}
\begin{equation}
\sqrt{\frac{\beta}{2\pi}} e^{-\beta
  X_2^2/2}=Q_2(X_2)\int\limits_{-\infty}^\infty Q_1(X_1) e^{-\lambda
  X_1X_2} dX_1,
\end{equation}
where the sub- and super-indexes have been dropped for notational lightness. 
Substituting $Q_1(X_1)$ and $Q_2(X_2)$ by Gaussians, it is easy to see
that the required maximum entropy joint distribution is:
\begin{equation}
P_{(1,2)}(X_1,X_2)=\frac{1}{2\pi}\left(\frac{\alpha\beta}{1-\alpha\beta\Gamma^2}
\right)^{1/2} e^{-\frac{1}{2(1-\alpha\beta\Gamma^2)}[\alpha X_1^2+\beta
    X_2^2 + 2\alpha\beta\Gamma X_1X_2]}\label{jointG}
\end{equation}
Perhaps not unexpectedly, the maximum entropy joint distribution for
more variables with Gaussian marginals will be again a Gaussian
distribution with appropriate correlations. At this point it is worth
mentioning that if we restrict ourselves to the class of continuous
functions, then the set of joint distributions having prespecified
marginals and covariance is convex, and the concavity of the entropy
functional guarantees that the distribution at which it is maximized
is unique. However, a more difficult problem concerns whether
distributions satisfying the requirements exist at all (Note, for
example, that for large enough $\Gamma$ in eq.(\ref{jointG}), the
argument of the exponential changes sign in which case
$P_{(1,2)}(X_1,X_2)$ cannot be interpreted as a probability
distribution). Unfortunately, the general conditions under which the
set of distributions having the prescribed marginals and covariance is
not empty are not easy to establish \cite{ghosh}.  Finally, as upon
integration Gaussians beget Gaussians, for these distributions
eq. (\ref{2point3}), as well as generalizations to more variables,
will always hold.

Explicit expressions for maximum entropy joint
distributions corresponding to non-Gaussian marginal distributions
appear to be very hard to obtain. However, a perturbation expansion in
powers of the parameter $\lambda$ can be carried out rather
easily. Writing
\begin{equation}
Q_1(X_1)=P_1(X_1)(1+f_1^{(1)}(X_1) \lambda+f_1^{(2)}(X_1) \lambda^2+... )
\end{equation}
\begin{equation}
Q_2(X_2)=P_2(X_2)(1+f_2^{(1)}(X_2) \lambda+f_2^{(2)}(X_2) \lambda^2+... )
\end{equation}
in equations (\ref{Coupledeqs1}) and grouping powers of $\lambda$,
after some rather messy algebra, one can write that, correct to order
$\lambda^2$
\begin{eqnarray}
&&P_{(1,2)}(X_1,X_2)=P_1(X_1)P_2(X_2) \times \cr
&&e^{-\left\{\lambda(X_1-\langle X_1\rangle )(X_2-\langle X_2\rangle )
 +\frac{\lambda^2}{2} \left((\langle X_2^2\rangle -\langle X_2\rangle ^2)(X_1-\langle X_1\rangle )^2+
  (\langle X_1^2\rangle -\langle X_1\rangle ^2)(X_2-\langle X_2\rangle )^2- (\langle X_1^2\rangle -\langle X_1\rangle ^2)(\langle X_2^2\rangle -\langle X_2\rangle ^2)\right)\right\}} 
\nonumber \label{pert}
\end{eqnarray}
where the averages, $\langle X_1\rangle $, $\langle X_1^2\rangle $, etc., are taken over the
marginal distributions, assuming the moments exist, and $\lambda$ is
calculated using equation (\ref{lambda1}):
\begin{equation}
\lambda\approx \frac{\langle X_1X_2\rangle -\langle X_1\rangle \langle X_2\rangle }{(\langle X_1^2\rangle -\langle X_1\rangle ^2)(\langle X_2^2\rangle -\langle X_2\rangle ^2)}+...
\end{equation}
Clearly, writing the joint distribution as an exponential in
eq.(\ref{pert}) is not really warranted, except by the fact that it
guarantees both positivity and integrability, and that it turns out to
be slightly more compact than might have been expected. Further, it
also highlights the fact that the approximation for
$P_{(1,2)}(X_1,X_2)$ has the form required by eq. (\ref{Form}),
corresponding to a maximum entropy distribution. From this expression,
approximate conditional distributions can be derived immediately, as
well as conditional expectations. Thus, for example, the conditional
expectation of $X_1$ given $X_2$, to linear order in $\lambda$, is
\begin{eqnarray}
\langle X_1|X_2\rangle  &\approx& \langle X_1\rangle  -\lambda (\langle X_1^2\rangle -\langle X_1\rangle ^2)(X_2-\langle X_2\rangle ) + ... \cr
&&\cr
&=& \langle X_1\rangle  -\frac{(\langle X_1X_2\rangle -\langle X_1\rangle \langle X_2\rangle )(X_2-\langle X_2\rangle )}{\langle X_2^2\rangle -\langle X_2\rangle ^2}+...
\end{eqnarray}
Also, the excess entropy over the product of marginals is found to be
\begin{eqnarray}
&&\Delta S=\int\limits_{I_1\times I_2} P_{(1,2)}(X_1,X_2)\ln
\left(\frac{P_{(1,2)}(X_1,X_2)}{P_1(X_1)P_2(X_2)}\right)dX_1 dX_2 \approx \cr
&&-\lambda^2(\langle X_1^2\rangle -\langle X_1\rangle ^2)(\langle X_2^2\rangle -\langle X_2\rangle ^2)+... = -\frac{(\langle X_1X_2\rangle -
\langle X_1\rangle \langle X_2\rangle )^2}{(\langle X_1^2\rangle -\langle X_1\rangle ^2)(\langle X_2^2\rangle -\langle X_2\rangle ^2)}+... \nonumber
\end{eqnarray}

The three point distribution can be obtained in the same way, but the
result is too long and unenlightening to include here. Nevertheless,
it should be mentioned that at least to second order in the
perturbation parameter, equation (\ref{2point3}) still holds (assuming
that all the correlation coefficients can be considered to be of
linear order in the perturbation parameter).

In summary, the construction of maximum entropy joint probability
distributions with the prescribed marginals and covariance has been
discussed. It should be noted that while there are other convenient
methods for constructing joint probability distributions with the
prescribed marginals and covariance \cite{ghosh}, only when the
entropy is maximized can we be sure that no extra, uncontrolled
assumptions have been introduced.

Extensions to even more variables are straight forward in principle,
but the set of coupled equations that result from the maximization of
entropy is larger and harder to solve. The exception being, as
mentioned earlier, the case of gaussian marginals with fixed
correlations, for which the maximum entropy distribution is the
appropriate correlated gaussian distribution. Also, the whole
discussion can be extended to the case in which the inter-relation
among the variables is not encoded in the linear correlation constant,
but rather by another more general moments; for example, for the case
of random variables $X_1$ and $X_2$ with given marginals, with the
constriction $\langle f(X_1,X_2)\rangle =0$. Another interesting
extension pertains to approximation schemes for the joint
distribution. For example, for the case in which the marginals do not
have second moment, so the perturbation expansion as presented above,
is not possible.

I am grateful to Ana Mar\'ia Contreras for her valuable comments on the
manuscript, and to F. Leyvraz and S. Majumdar for useful discussions.
Partial support through grant DGAPA-UNAM IN109111 is gratefully
acknowledged.


\begin{thebibliography}{00}

\bibitem{Bouchaud} ``Theory of Financial Risk and Derivative Pricing
  From Statistical Physics to Risk Management'' Second Edition.
  J.-P. Bouchaud and M.Potters Cambridge University Press 2003

\bibitem{Markus} The influence of static correlations on multivariate
  correlation analysis of the EEG C. Rummel G. Baier, M. Müller
  Journal of Neuroscience Methods {\bf 166}, 138–157 (2007)

\bibitem{Schindler} Brain (2007), {\bf 130}, 65-77 Assessing seizure dynamics
  by analysing the correlation structure of multichannel intracranial
  EEG K. Schindler, H. Leung, C.E. Elger and K. Lehnertz

\bibitem{Nicolis} G. Nicolis G and M.M. Mansour ``Onset of spatial correlations
  in nonequilibrium systems: a master-equation description''
  Phys. Rev. A {\bf 29} 2845 (1984)

\bibitem{Dorfman} Dorfman J R, Kirkpatrick T R and Sengers J V 1994 ``Generic
  long-range correlations in molecular fluids''
  Ann. Rev. Phys. Chem. {\bf 45} 213 (1994)

\bibitem{Zarate} J.O.M. de Zarate and J.V. Sengers 2006 ``Hydrodynamic
  Fluctuations in Fluids and Fluid Mixtures'' (Amsterdam: Elsevier)

\bibitem{HL} ``Long-range correlations in a simple stochastic model of
  coupled transport'' H. Larralde and D. P. Sanders 2009 J. Phys. A:
  Math. Theor. {\bf 42} 335002

\bibitem{Schmidt} Schmidt, T. ``Coping with Copulas''. In Copulas - From
  Theory to Application in Finance, J. Rank (ed.), 3-34, Risk Books,
  London (2007)

\bibitem{Genest} The Joy of Copulas: Bivariate Distributions with Uniform
  Marginals C. Genest and J. MacKay The American Statistician, {\bf
    40}, No. 4 (Nov., 1986), pp. 280-283

\bibitem{Piant1} J. Piantadosi, P. Howlett and J. Boland JOURNAL OF
  INDUSTRIAL AND MANAGEMENT OPTIMIZATION Volume 3, Number 2,
  pp. 305-312, (2007)

\bibitem{Piant2} Optim. Lett. (2012) {\bf 6} 99-125 Copulas with maximum
  entropy J. Piantadosi P. Howlett J. Borwein

\bibitem{granger} C.W.J.Granger, Econometrica {\bf 37}, 424-438, (1969).

\bibitem{Jaynes} E.T. Jaynes; Phys. Rev. {\bf 106} 620-630, (1957).

\bibitem{ghosh} ``DEPENDENCE IN STOCHASTIC SIMULATION
MODELS'' Soumyadip Ghosh, PhD. Thesis (2004); available at
http://people.orie.cornell.edu/shane/theses/GhoshThesis.pdf

\end{thebibliography}
\end{document}